\documentclass[letterpaper, 10 pt, conference]{ieeeconf}  

\IEEEoverridecommandlockouts
\usepackage{cite}
\usepackage{amsmath,amssymb,amsfonts}
\usepackage{algorithmic}
\usepackage{graphicx}
\usepackage{textcomp}
\usepackage{xcolor}
\def\BibTeX{{\rm B\kern-.05em{\sc i\kern-.025em b}\kern-.08em
    T\kern-.1667em\lower.6ex\hbox{E}\kern-.125emX}}
    
{}
{}
{}
\usepackage{caption}
\usepackage{subcaption}
\usepackage[acronym]{glossaries}
\glsdisablehyper
\DeclareMathOperator{\atan2}{atan2}
\usepackage{hyperref}
\usepackage{booktabs}

\newacronym{ev}{EV}{Electric Vehicle}
\newacronym{adas}{ADAS}{advanced driver-assistance systems}
\newacronym{ipa}{IPA}{Intelligent Personal Assistant}
\newacronym{iva}{IVA}{Intelligent Vehicle Assistant}

\newacronym{dsp}{DSP}{Digital Signal Processing}
\newacronym{can}{CAN}{Controller Area Network}
\newacronym{ui}{UI}{User Interface}
\newacronym{avp}{AVP}{automated valet parking}
\newacronym{iot}{IoT}{internet of things}

\newacronym{ux}{UX}{user experience}
\newacronym{ueq}{UEQ}{User Experience Questionnaire}
\newacronym{ecu}{ECU}{Electronic Control Unit}

\usepackage{pgfplots}
\usepackage{tikz}
\usetikzlibrary{positioning}

\usepackage{color, colortbl}
\definecolor{Gray}{gray}{0.9}

\usepackage{longtable}

\usepackage{array}

\let\labelindent\relax
\usepackage{enumitem}

\usepackage[tikz]{ocgx2}
\usepackage{pgfplotstable}
\pgfplotsset{width=12.2cm, height=7cm, compat=1.14}
\usepackage{svg}

\usepackage[font={footnotesize}]{caption}

\usepackage{amssymb}
\usepackage{booktabs,xltabular}

\begin{document}

\title{\LARGE \bf Park4U Mate: Context-Aware Digital Assistant for Personalized Autonomous Parking\\
}

\author{
{Antonyo Musabini$^{1}$, Evin Bozbayir$^{2}$, Hervé Marcasuzaa$^{1}$, Omar Adair Islas Ramírez$^{1}$} \\
{Valeo, Comfort \& Driving Assistance Systems, Innovation \& Collaborative Research} \\
{$^{1}$93000 Bobigny, France, $^{2}$74321 Bietigheim-Bissingen, Germany} \\
{\tt\small \{antonyo.musabini, evin.bozbayir, herve.marcasuzaa\}@valeo.com}, {\tt\small omar.islasrm@gmail.com}%
}

\maketitle
\thispagestyle{plain}
\pagestyle{plain}

\begin{abstract}
People park their vehicle depending on interior and exterior contexts. They do it naturally, even unconsciously. For instance, with a baby seat on the rear, the driver might leave more space on one side to be able to get the baby out easily; or when grocery shopping, s/he may position the vehicle for trunk accessibility. Autonomous vehicles are becoming technically effective at driving from A to B and parking in a proper spot, with a default way. However, in order to satisfy users’ expectations and to become trustworthy, they will also need to park or make a temporary stop, appropriate to the given situation. In addition, users want to understand better the capabilities of their driving assistance features, such as automated parking systems. A voice-based interface can help with this and even ease the adoption of these features. Therefore, we developed a voice-based in-car assistant (\textit{Park4U Mate}), that is aware of interior and exterior contexts (thanks to a variety of sensors), and that is able to park autonomously in a smart way (with a constraints minimization strategy). The solution was demonstrated to thirty-five users in test-drives and their feedback was collected on the system’s decision-making capability as well as on the human-machine-interaction. The results show that: (1) the proposed optimization algorithm is efficient at deciding the best parking strategy; hence, autonomous vehicles can adopt it; (2) a voice-based digital assistant for autonomous parking is perceived as a clear and effective interaction method. However, the interaction speed remained the most important criterion for users. In addition, they clearly wish not to be limited on only voice-interaction, to use the automated parking function and rather appreciate a multi-modal interaction.

\end{abstract}


\section{Introduction} \label{section:Introduction}
\thispagestyle{empty}

Parking is known to be a stressful part of the driving experience. Drivers often become nervous while searching a suitable parking spot. Maneuvering into that spot might be an anxious task for users new to driving as well as users new to a car model.

The automotive industry tackles this issue with dedicated \gls{adas}. For instance, Tesla vehicles include an automated parking solution called AutoPark~\cite{TeslaManual}, where the driver should manually drive in front of a parking lot. Once the vehicle detects the spot, it can park in autonomously. However, using automated parking \gls{adas} without knowing its limitations can lead to a high degree of initial distrust~\cite{Tenhundfeld2020}. The absence of efficient feedback on sensor data and decision process of the vehicle leads the users to judge the vehicle's decision-making process inefficient or unadapted to the current context. Therefore, the usability of current autonomous parking systems are still weak. 72\%~of drivers don't trust active parking-assist systems~\cite{AAA2015}.

While selecting a parking strategy, the driver and occupants take the following points (and more) into consideration, to decide where and how to park their vehicle:
\begin{itemize}
    \item Journey (e.g., Are we near to the destination? Is it for a shortstop? for stationing all night?);
    \item Exterior (e.g., Is it raining? Is there traffic on the road?);
    \item Interior (e.g., Where are the occupants sitting? Can they get out easily? Are there luggage in the trunk?);
    \item Habits (e.g., How does the driver usually park at that location?);
    \item Preferences (e.g., How do the occupants would like to be parked in that circumstance?)
\end{itemize}

Being able to take into consideration these kinds of parameters becomes essential for autonomous vehicles, like those of a robo--taxi service, as vehicles must be able to park intelligently for the needs of the passengers (at pick up or drop off) and the vehicles may not have any control commands any longer. Whereas, today, research on autonomous vehicles concentrates deeply to the travel time-related features~\cite{Tenhundfeld2020}. Shuttles and robo--taxis need to decide how to park with the knowledge of the current context. This capability would also be a great comfort and a safety \gls{adas} feature for standard vehicles.

Once drivers understand the systems' operation and decision-making, they are much more likely to trust it, which increases the technological acceptance~\cite{Tenhundfeld2020}. Our goal is therefore to improve automated parking by proposing a personalized computational model that detects the most suitable parking spot and decides how to position the vehicle in that space, adapting to the interior of the vehicle (who sits where in the car?) and the external context (is this a suitable parking space?). Despite the context-awareness, a natural, voice based interaction between the driver and the car is another key element of \textit{Park4U Mate}, helping the driver to stay informed in the whole parking process, from searching for a parking spot to maneuvering until ending the parking maneuver.

Section~\ref{section:SoA} details the state-of-the-art human-machine interfaces dedicated to automated parking and the global use of in-vehicle digital assistants; section~\ref{section:Methodology}, explains the vehicle sensors and the experimental protocol (mainly the dialogues and the parking adaptation algorithm). Then, section~\ref{section:Results} presents the obtained results in terms of user experience and acceptability, and finally section~\ref{section:Conclusion} presents the conclusion.

\section{Related Work}\label{section:SoA}
\thispagestyle{empty}

\subsection{Existing Assistants and Parking Systems}

Human-computer cooperation based features for parking has various autonomy levels. While in Tesla's AutoPark feature the driver should manually drive in front of a parking spot before the autonomous maneuver execution~\cite{TeslaManual}, other features propose complete autonomous solution.  For example, Valeo's Park4U Home feature is an \gls{avp} solution that allows the vehicle to park autonomously in a parking spot without the need for a human to be present in the vehicle, provided the vehicle is trained by the driver on how to park in the home's parking spot~\cite{Park4UHome}. In order to overcome the predefined (or trained) parking spot limitation of \gls{avp}, within the AutoPilot project, a mini drone cooperates with the vehicle to identify vacant parking spots nearby~\cite{Tcheumadjeu2018}. Unfortunately, despite the obvious advantages of \gls{avp}, legal regulation in terms of human absence in an autonomous vehicle is still a major obstacle to this feature. Moreover, none of the existing technology is able to adapt how the vehicle should park in terms of the current circumstances. Therefore, the existing systems should communicate better with the end users and be more context-aware.~\textit{"Once drivers understood what the system was doing, they were much more likely to trust it"}~\cite{Tenhundfeld2020}.

Recently, digital assistants such as Amazon Alexa, Google Now, Apple Siri and Microsoft Cortana offer straightforward implementation possibilities for custom chatbots~\cite{larsson2017user}. Hence, voice-based human-machine interactions are emerged for various use cases. More specifically in some high-end commercial vehicles, current assistants are implemented for some command based interactions such as controlling the air conditioner or setting a navigation point (i.e., MBUX - Mercedes-Benz User Experience feature for A-Class vehicles~\cite{CerenceMercedes}). However, these assistants are not able to communicate with sensor data and adapt their proposals. The Dragon Drive Framework of Cerence proposes to reduce the in-car and outside-the-car gap by enchaining the digital assistant with the driver's gaze information and landmarks nearby~\cite{villemure2018dragon}. Similarly to this technology, Gestigon proposes to reduce this gap with a gesture detection sensor (the car knows if the drivers point a specific landmark, then it takes into account while interacting)~\cite{Gestigon, coleca2013real}. Other solutions, such as CloudMade Digital Assistant Penalization Layer~\cite{CloudMade}, augments the current context with the driver's experience, knowledge, feature usage history, trip information and calendar information. With the aim of analyzing more in depth the current context, frameworks like Nuance Reasoning~\cite{jain2018nuance} and Paragon Semvox~\cite{ParagonSemvox}, gathers as much as external data and share the created sensor augmented context with other features of the vehicle. Finally, thanks to this kind of complete context, beyond sensor augmentation and contextualization, more advance assistants tend to create an emotional engagement with drivers and passengers (i.e., NIO’s digital assistant, called NOMI~\cite{NIO_NOMI} or BMW Intelligent Personal Assistant~\cite{bmw}).

In light of the presented details, we propose an \gls{adas} feature, on the top of Valeo's Park4U system~\cite{valeo_park4u} (an automated parking system, already available in commercial vehicles). The proposed feature is novel in two distinct points. First, with a very wide interior and exterior sensor panel, we collect data of the current situation and analyze the context. This analysis computes a best matching scoring between available parking spots and different parking possibilities (i.e., backwards, forwards, close to the front vehicle, away from the road). Second, a voice-based digital assistant handles the interaction between the driver and the vehicle. We assume that the way of interaction, the adaptive parking and provided information on the parking process will accelerate the acceptance of the automated parking features, especially to the novice users. It will also ease the system activation interaction, increase transparency in systems' capabilities and increasing trust in automated systems. According to our research, this is the first voice interactive parking assistant able to park autonomously in a user-friendly and context-aware way. Accordingly, this is also the first user experience study on such a system.

\subsection{Preliminary User Study} \label{subsection:UserStudies}

The idea to create an interactive voice-based digital assistant for automated parking came up in an initial study conducted in 2019. The purpose was to understand users' expectations and needs on automated parking systems and to translate the findings into solutions to improve the parking experience. The study participants, 4 female and 8 male between 25 and 65 years, were all first-time users of automated parking systems. First time users were perceived as relevant target group, as the intention was to develop intuitive interaction solutions that don't require any prerequisite knowledge. As a key stimulus, participants tested the fully automated parking system of a commercial vehicle 
from 2016. The main research method in this study was to first let users discover and test the system blindly, before they got a tutorial on how to activate and use the parking function.
The preliminary study revealed that users had difficulties with the commercial product in all three stages of the automated parking process: searching for a parking spot, deciding for a suitable parking spot and the parking maneuver itself. Besides the failing performance in some cases, one major frustration was the activation of the parking function and the interaction itself. Several steps needed to be completed until the parking maneuver actually started and display warnings failed their intention to engage users to keep their eyes on the road. Besides the fact that some users mentioned voice based interaction as the ideal way of activating the parking process, the study helped to understand how voice interaction could also overcome the issue of users' knowledge gap in the systems capabilities and limitations while engaging a self-explanatory and natural interaction. This way of interaction in turn is one key feature of \textit{Park4U Mate}, which was again evaluated by users and which will be described in the following. 

\section{Methodology}\label{section:Methodology}
\thispagestyle{empty}

\subsection{User Study on Park4U Mate}\label{subsection:User Study on Park4UMate}

\subsubsection{User Panel}
In total, 35 users experienced and evaluated Park4U mate, which can be divided into two subgroups: One group consisted of 19 research and development professionals (18 male and 1 female) of automated parking systems; another consisted of 16 users (8 male and 8 female) with no professional background in automated parking, referred to as "novices." This selection ensured a diversity of perspectives to evaluate the solution.

\subsubsection{Pre-test Interviews} \label{PreTestInterviews}
All users were taking part in a short pre-test interview before the actual \textit{Park4U Mate} system was demonstrated. The aim was to understand their general definitions of assistants and their current parking habits before the possible bias in their response after seeing the demonstration. The following questions were asked to the study participants: "\textit{How do you define an assistant?}", "\textit{What would an assistant mean to you?}", "\textit{How do you choose a parking space?}", "\textit{Are there certain parameters that influence this choice?}", "\textit{If so, which ones?}". 

\subsubsection{Evaluation Methods} \label{EvaluationMethods}
Qualitative as well as quantitative data was collected by both user groups during and after experiencing the \textit{Park4U Mate} system in test drives.

\begin{figure}[!htb]
    \centering
    \includegraphics[width=.445\linewidth]{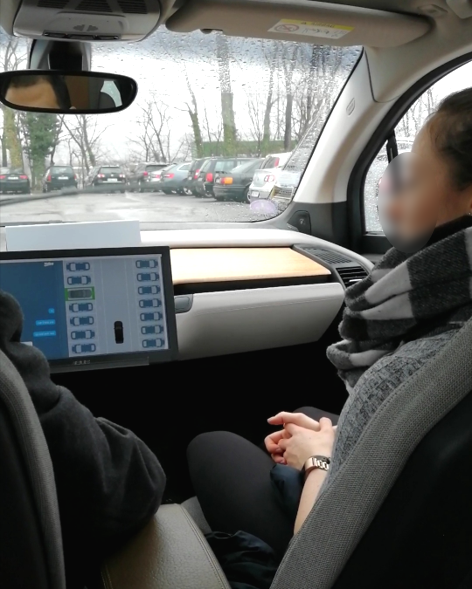}
    \caption{An image taken while the user studies. A participant is observing the interactions as co-driver. The user interface is visible at the center of the vehicle. It shows the dialogues on left and the current parking area as well as a parking proposition on the right side of the screen.}
    \label{fig:user_test_i3}
\end{figure}

Qualitative data consisted of users' general thoughts and reactions. During test drives, the system was demonstrated by a test driver and users were observing the interactions as co-drivers (see Figure \ref{fig:user_test_i3}). A researcher sitting in the backseat asked users to think out loud what they saw and felt, and observed their reactions. They were also asked to express what they liked and disliked about the solution. The users were furthermore requested not to evaluate the automated parking system itself, but to try to focus on the voice-based digital assistant and its decision-making process. All the manoeuvres were conducted in a private parking area.

Quantitative data was collected with a post-test: \gls{ueq}~\cite{HINDERKS201938}. The questionnaire covers a comprehensive impression on both classical usability aspects and user experience aspects. The six investigated points are explained as follows: \textbf{Attractiveness}: \textit{"Overall impression of the product. Do users like or dislike it?"}, \textbf{Perspicuity}: \textit{"Is it easy to get familiar with the product and to learn how to use it?"}, \textbf{Efficiency}: \textit{"Can users solve their tasks without unnecessary effort? Does it react fast?"}, \textbf{Dependability}: \textit{"Does the user feel in control of the interaction? Is it secure and predictable?"} \textbf{Stimulation}: \textit{"Is it exciting and motivating to use the product? Is it fun to use?"} and \textbf{Novelty}: \textit{"Is the design of the product creative? Does it catch the interest of users?"}~\cite{UEQ}. 24 out of 35 users returned the post-questionnaire filled (12 responses from novice and 12 from expert users).

\subsection{Equipment} \label{subsection:Materials}
\thispagestyle{empty}

We used a commercial \gls{ev}, with autonomous parking functions, that we modified with the following set of equipment (see Figure~\ref{fig:i3_seonsors}):

\begin{figure}[!htb]
    \centering
    \includegraphics[width=.89\linewidth]{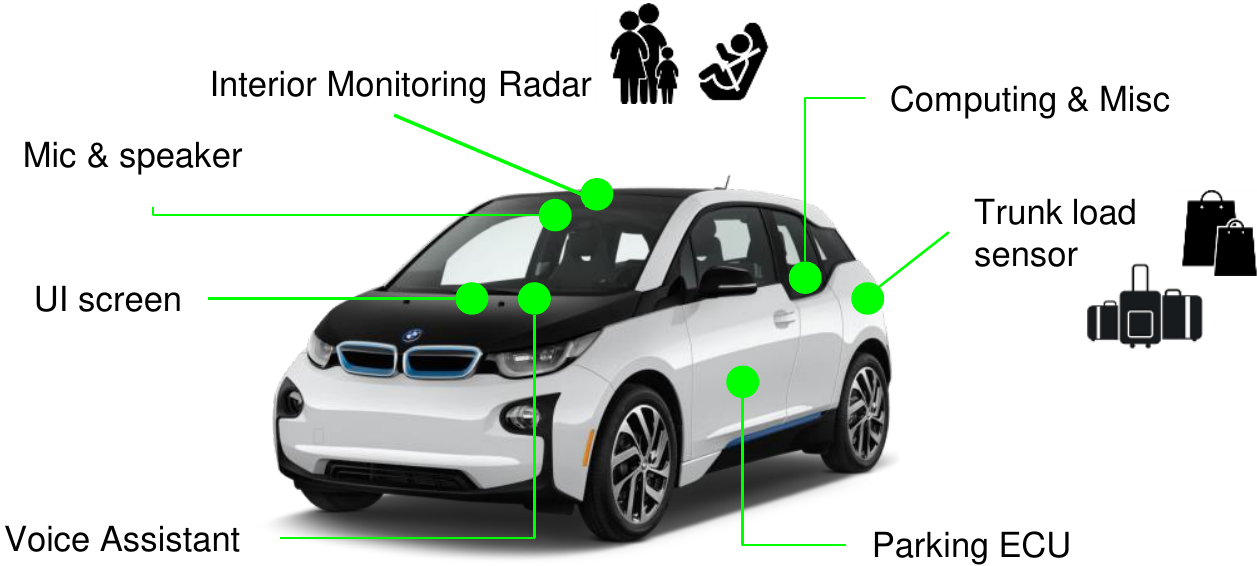}
    \caption{The experimentation vehicle with the additional sensor panel. The green circles represent where the sensors are located into the vehicle.}
    \label{fig:i3_seonsors}
\end{figure}

\begin{itemize}
    \item A \textit{\gls{ui}} screen was placed on the central cluster, where a tailor-made application was displayed (see Section~\ref{subsection:UserInterface} for details).
    \item A \textit{radar} was placed on the roof of the vehicle to detect seat occupancy and occupant types (adult or baby).
    \item An ultrasonic \textit{telemeter} was placed into the trunk to detect if it is full or not.
    \item A pair of \textit{microphones} and a \textit{speaker} were installed next to the top interior light. These sensors were connected to a noise canceling \gls{dsp} unit.
    \item A Park4U \textit{\gls{ecu}}, able to receive parking commands from CAN--Bus, was installed to the vehicle. The commercial Park4U product was modified for the purpose of the prototype. It was also sharing the ongoing autonomous parking manoeuvre's status.
    \item A \textit{4G cellular connection} network was installed to be able to communicate with the cloud services (i.e., text to speech, speech to text).
    \item A \textit{GPS antenna} was installed to the vehicle, needed for proactive test scenarios (i.e., the vehicle understood that it was in a parking area).
\end{itemize}
They were all connected to an automotive \textit{computer} in the trunk. Most data transfers were carried over MQTT and the software was developed in Python.

\subsection{The User Interface}\label{subsection:UserInterface}

\begin{figure}[!htb]
    \centering
    \subfloat[\centering Bird--view parking area view]{\includegraphics[width=.445\linewidth]{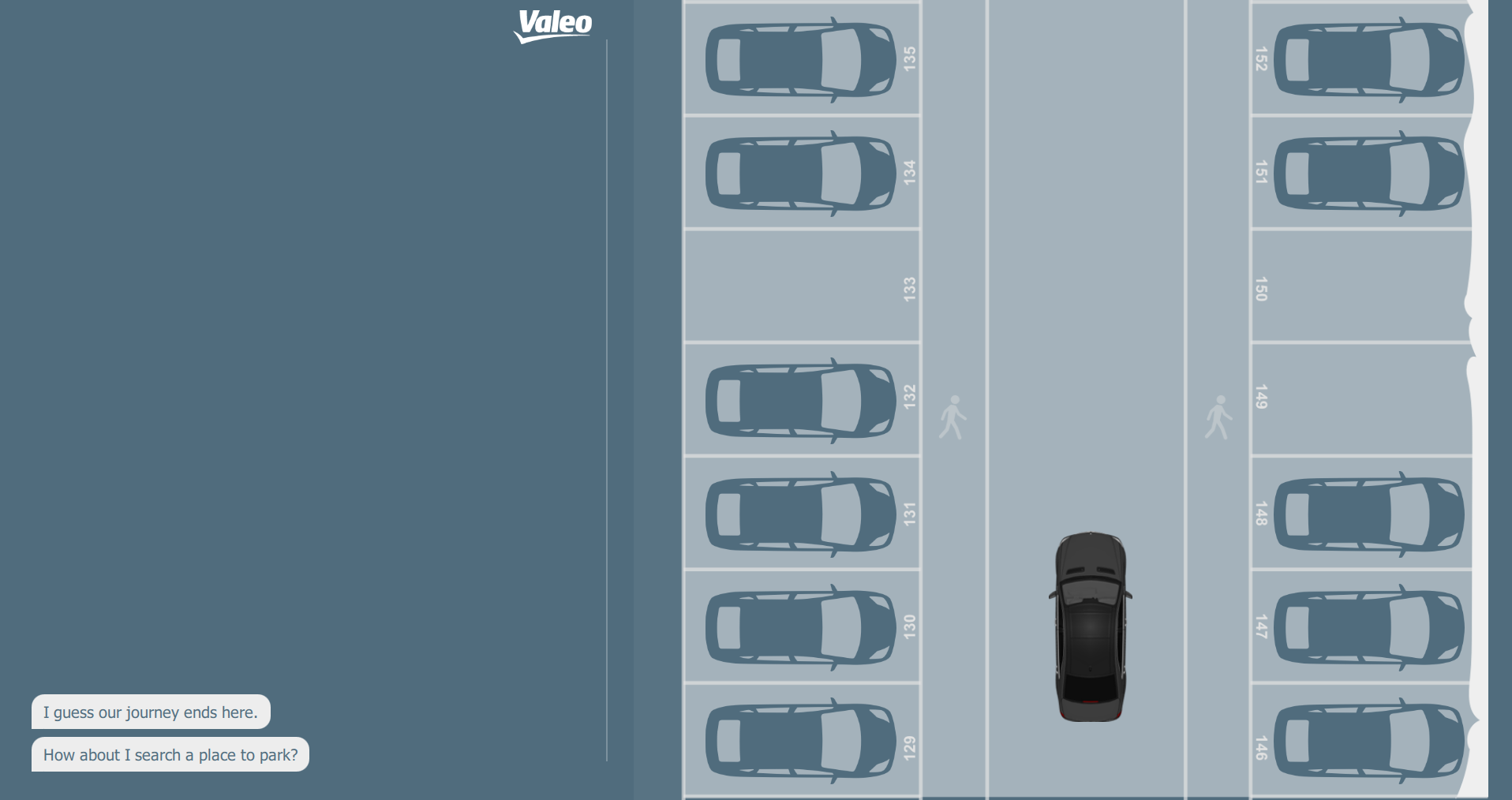}\label{figure:hmi:proactive} }
    \qquad
    \subfloat[\centering Cockpit view]{\includegraphics[width=.445\linewidth]{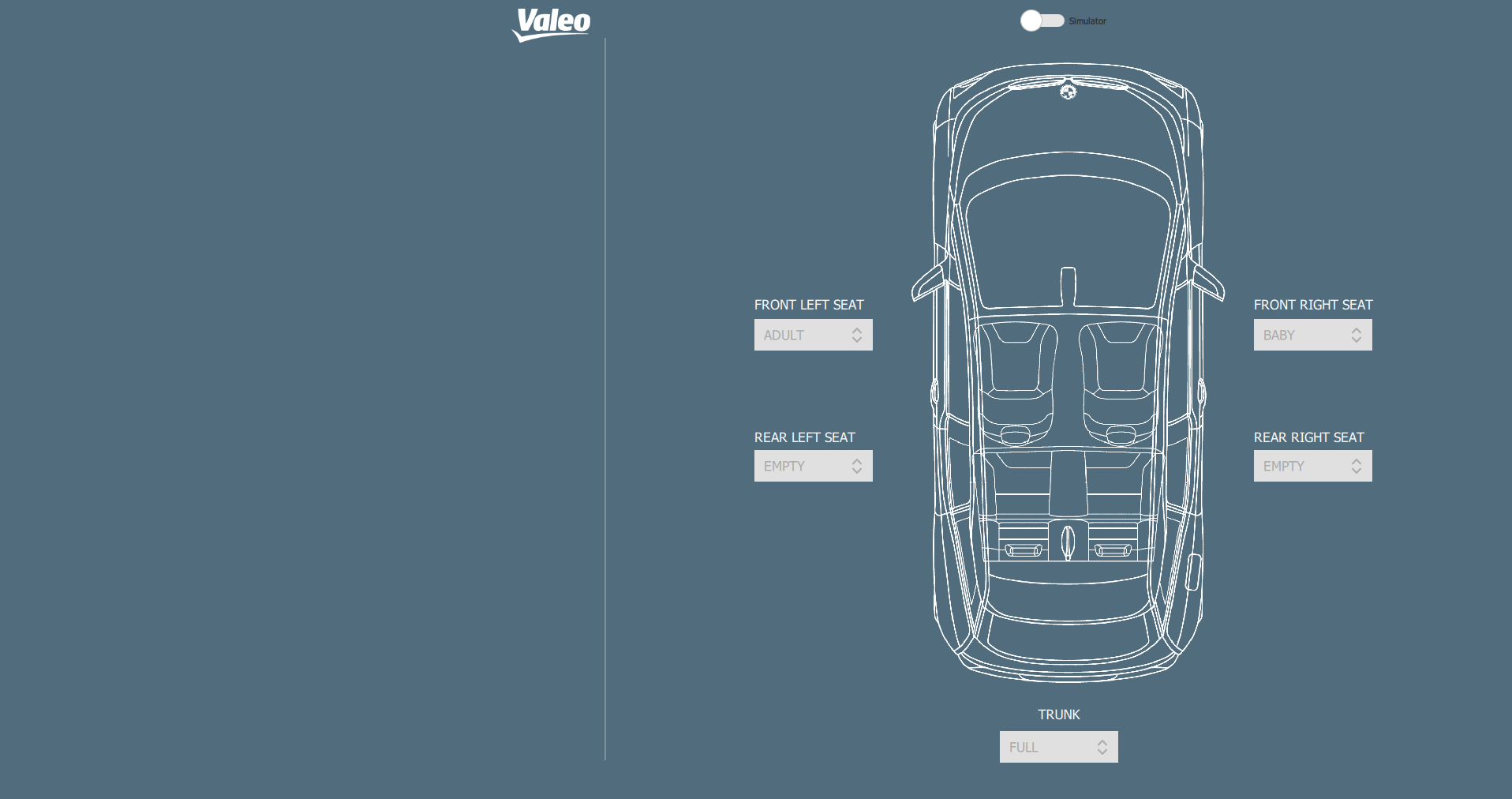}\label{figure:hmi:Cockpit} }
    \qquad
    \subfloat[\centering Proposition view]{\includegraphics[width=.445\linewidth]{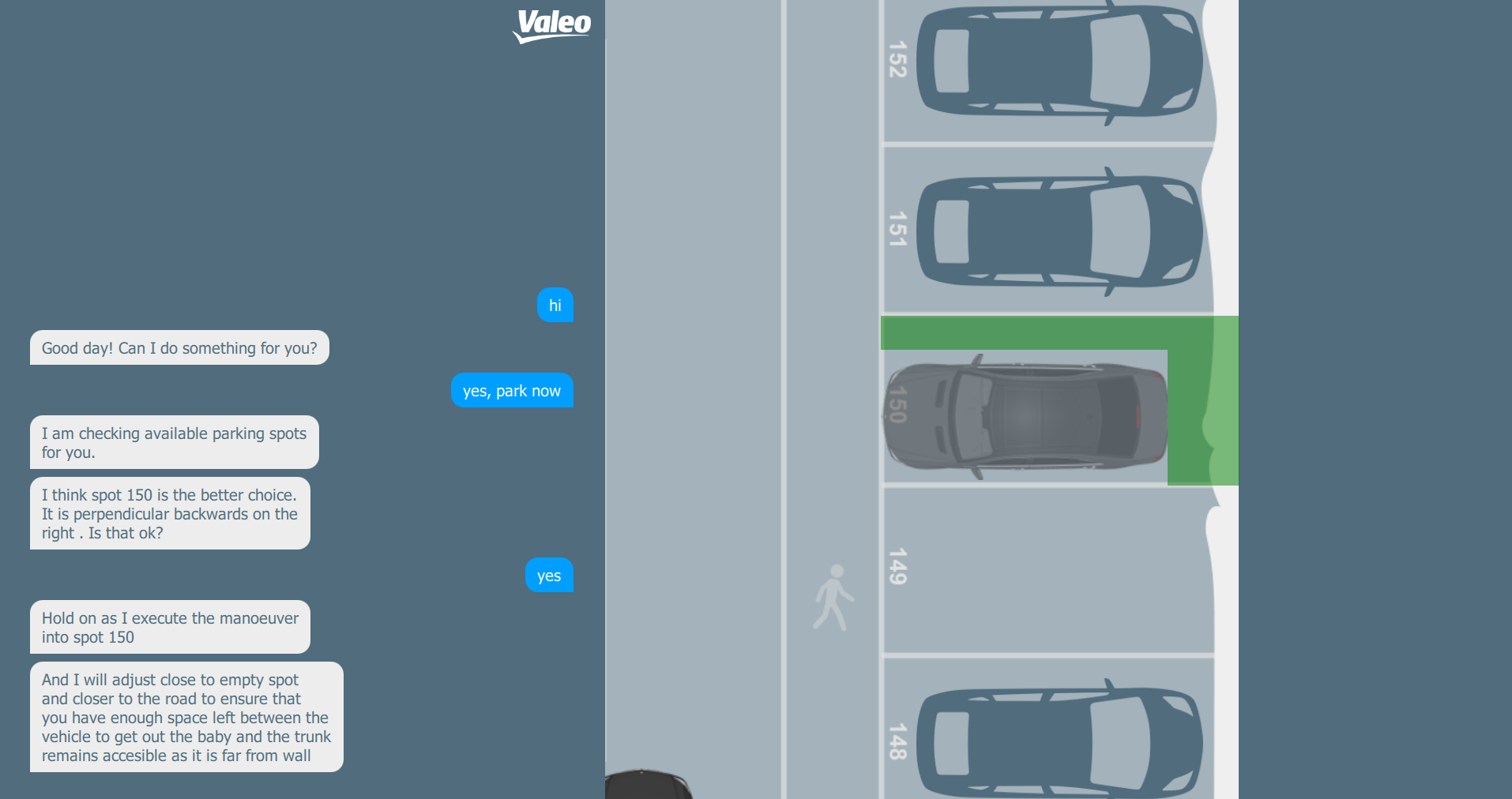}\label{figure:hmi:proposition} }
    \caption{The three views of the \gls{ui}. a) The parking area with three vacant places is visible on the right. The ego-vehicle is illustrated as a black car. Some proactive dialogues are present on the left. b) The occupants and the trunk status is shown on right. On left there is a dialogue view, but there is no any dialogue yet. c) After some dialogues, the vehicle proposes the best suited parking strategy and the interface zooms on that place. The final position of the parking strategy is illustrated as a transparent ego vehicle with green areas left around for illustrate available space maximization. Here, the vehicle proposes to park on a perpendicular spot, backwards, with maximizing remaining space on the left and at the back.}
    \label{figure:HMI}
\end{figure}

A two-parts \gls{ui} was developed for the central cluster (see Figure~\ref{fig:user_test_i3}). On the left side, the ongoing driver-vehicle dialogue was shown, as seen in messaging applications . On the right side, the content was automatically switching between three views, adapting to the content of the dialogue. The first view displayed the parking area map, with each parking spot visible (see Figure~\ref{figure:hmi:proactive}). The second view represented of the cockpit, with info gathered from passengers and trunk sensors  (see Figure~\ref{figure:hmi:Cockpit}). The third view illustrated the parking strategy computed by the vehicle (see Figure \ref{figure:hmi:proposition}). The only way to interact with the \gls{ui} was voice; the driver could answer spontaneously to a question asked by the vehicle or a dialogue could be initiated by pronouncing "Hey Valeo" keyword.

\subsection{Digital Assistant and Dialogues}\label{subsection:DigitalAssistantandDialogues}
\thispagestyle{empty}

Gathered contextual information was linked with the assistant. Google's DialogFlow, speech-to-text and text-to-speech APIs were used. The dialogue is shaped by triggers, clustered under two titles: \textit{intents} and \textit{events}. An \textit{intent} is the understanding of what driver pronounced, by the natural language processing tool, and an \textit{event} is based on what physically happened (i.e., sensor data). Table~\ref{table:dialogueflow:intents} illustrates the implemented intents and Table~\ref{table:dialogueflow:events} details the interpretable events.

\begin{table}[!htb]
\caption{Understandable intents from the digital assistant.\label{table:dialogueflow:intents}}
{\begin{tabular}{@{\extracolsep{\fill}}@{} p{2cm}|p{6.1cm} @{}}\toprule
Intent Type & Description \\\midrule
Welcoming                  & Welcoming words pronounced by the driver (i.e., hi, how are you) are in this category. The vehicle welcomes the driver as in return and asks how it can help.\\
\hline
Asking car representation       & The driver might ask the vehicle to cite the seat occupancy status or the trunk status for demo purposes.       \\    \hline 

Asking to park                  & Driver asks to the vehicle to park (i.e. \textit{"park now"}). If there is at least one available parking spot, the assistant analyzes the best parking option and proposes it by precising why this option is the most adapted one.       \\
\hline
Responding to a parking proposition                  & The driver might accept or refuse the proposed parking spot. If the driver refuses a parking spot and if there is another one available, the driver can choose it. \\
\hline
Insulting the car               &  Insults pronounced by the driver is invited to be polite.  \\
\hline
Thanking the car               & When the driver thanks the vehicle, it responds humbly. \\\bottomrule
\end{tabular}}{}
\end{table}

\begin{table}[!htb]
\caption{Interpretable events about the changes in the real world\label{table:dialogueflow:events}}
{\begin{tabular}{@{\extracolsep{\fill}}@{} p{2cm}|p{6.1cm} @{}}\toprule
Event Type & Description \\\midrule
At the destination                  & The vehicle arrives to the parking area and the assistant asks proactively if the driver wants to park in (based on GPS data).\\
\hline
Occupants change                  & One of the occupants in the car changes place or gets out or a new one gets in.\\
\hline
Trunk situation change                  & The trunk is loaded or unloaded. \\
\hline
Parking spot occupancy change                  & If a previously vacant parking spot becomes occupied or if a new one becomes vacant. \\
\hline
Park4U status   & States of the ongoing autonomous parking maneuver:~analyzing the spot, the spot is validated the maneuver will start, steering to target (forwards or backwards), failed / aborting, end of the maneuver.
\\\bottomrule
\end{tabular}}{}
\end{table}

\subsection{Parking Area Status Detection} \label{subsection:ParkingAreaStatusDetection}

The parking area and the vacant parking spot detection is the only simulated sensor of the experiment. The \gls{ui} is set with a configuration file, to match exactly to the current parking area, with the correct number of vacant places and with the orientation of the parked vehicles (backwards / forwards, lateral / perpendicular). The users are told that this information is shared from the infrastructure of the parking to the vehicle.

\subsection{Algorithmic Core For Parking Strategy} \label{subsection:AlgorithmicCore}

With the aim of computing the best parking spot with an adapted alignment strategy, an environment representation was fused from sensors and a potential field like optimization algorithm was developed.

\begin{figure}[!htb]
    \centering
    \subfloat[\centering Cartoon representation]{\includegraphics[width=.445\linewidth]{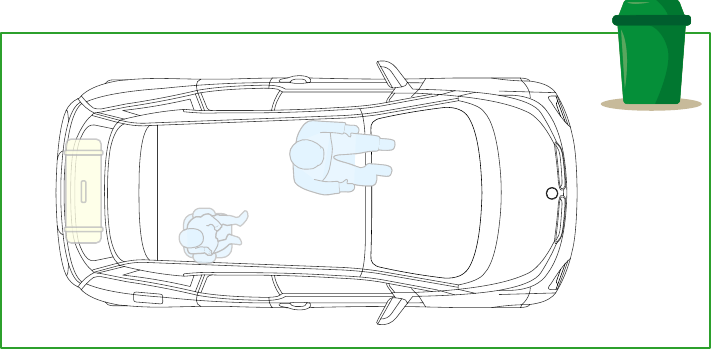}\label{fig:drawing_base} }
    \qquad
    \subfloat[\centering Geometric representation
    ]{\includegraphics[width=.445\linewidth]{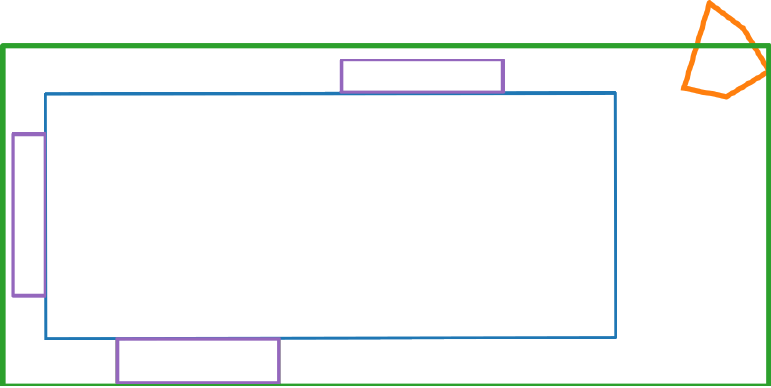}\label{fig:geometric_base_c} }
    \caption{Representations of the car and its environment. a) A scenario showing a driver, a baby and a luggage in the car. Also a garbage can is in one corner of the spot. b) Parking spot and the representation of the garbage can from \ref{fig:drawing_base} and the vehicle with needed extra space for occupants and the trunk.
    }
    \label{fig:main_representation}
\end{figure}

\subsubsection{Environment Representation}
The geometrical representation of the environment is illustrated as in Figure~\ref{fig:main_representation},
where the \textit{parking spot} is presented as a green rectangle, the \textit{car} is presented by a blue rectangle, \textit{passengers} and the luggage in the trunk are presented by rectangles of various sizes, depicting the degree of maneuvers needed to get out. Their sizes are in proportion with the user's specific parking habits and preferences. \textit{Obstacles} are presented by polygons that are going to obstruct some area of a parking spot. 

During demonstrations, the sizes of rectangles were fixed, but in a long term usage, the system should adapt itself depending on the driver's previous choices, also only the vehicles parked next to the empty spot were presented as obstacles (like in Figure~\ref{fig:result_5}).

\subsubsection{Potential Fields Extension}

The potential field extension of a parking is presented as successive polygons of elements $p_i \in P$, where $p_i=(x_i,y_i)$. This definition is extended as to have $p_i=(x_i,y_i,\alpha_i)$, where $\alpha_{i}=\atan2(y_{i+1}-y_i, x_{i+1}-x_i)$, where $i+1=0$ for the last element in order to form a cycle.

Each line of a polygon is presented as (\ref{eq:line_representation}).

\begin{equation}
  \label{eq:line_representation}
  f_{ij}(x,y)=a_{ij}x+b_{ij}y+c_{ij}
\end{equation}

\begin{figure}[!htb]
    \centering
    \includegraphics[width=.445\linewidth]{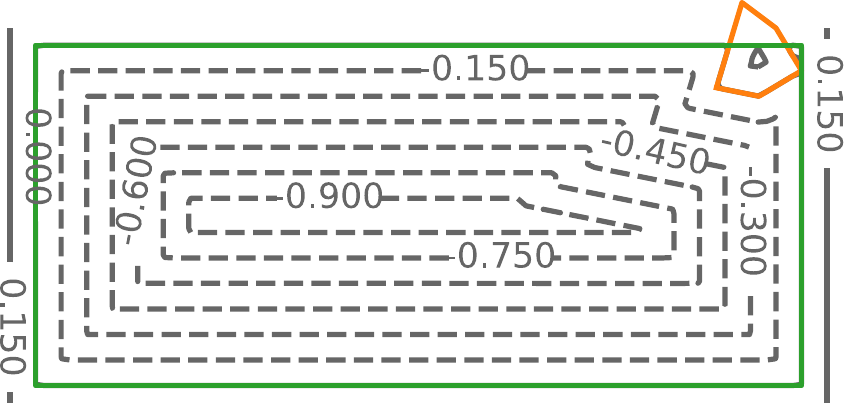}
    \caption{Example of force field generated by parking and a single object. The parking spot is illustrated as the green rectangle and the obstacle is colored orange. The generated force field is represented in gray doted lines, with the numbers corespondent to the height of the field.}
    \label{fig:force_field}
\end{figure}

Where $f_{ij}$ is the function of the $i$ polygon and its $j$ line. Where $a_{ij},b_{ij},c_{ij}$ are the parameters of the $ij$ line. Thus, $f_{ij}(x,y)$ is the distance/force value at location $(x,y)$ of line $ij$.

\begin{equation}
  \label{eq:space_representation}
  \Gamma (x,y) = \max_i\min_j f_{ij}(x,y)
\end{equation}
Where $\Gamma (x,y)$ is the force value of all polygons acting on $x,y$. The $min$ part of (\ref{eq:space_representation}) will get the force field effect of the $j$ polygon acting on position $(x,y)$ and the $max$ will select the most effective polygon at that position $(x,y)$. The example of force field generated by a single polygon is depicted in Figure~\ref{fig:force_field}.

Thus, since the force field is defined in (\ref{eq:space_representation}) and depicted in Figure~\ref{fig:force_field}, the area under the surface of the car rectangles (Figure~\ref{fig:geometric_base_c}) can be evaluated to compute the best parking strategy, within that parking spot.

As a parking area can have perpendicular, parallel or angled parking lots and the vehicle might need to park either forwards or backwards, the coordinate system needs to be transformed in order to unify them. This can be done by applying (\ref{eq:rotation}).
\begin{equation}
  \label{eq:rotation}
  \begin{bmatrix}
    \bar{x}
    \\
    \bar{y}
  \end{bmatrix}
  = \text{Rot}(\hat{\theta})
  \begin{bmatrix}
    x_l
    \\
    y_l
  \end{bmatrix}
  +
  \begin{bmatrix}
    \hat{x}
    \\
    \hat{y}
  \end{bmatrix}
\end{equation}

\begin{equation*}
    \text{Rot}(\hat{\theta}) = \begin{bmatrix} \cos \hat{\theta} & -\sin \hat{\theta} \\ \sin \hat{\theta} & \cos \hat{\theta} \end{bmatrix}
\end{equation*}

From~(\ref{eq:rotation}), $x_l$ and $y_l$ is the local coordinate system of the car which is then transformed to the global coordinate system by a rotation given the angle $\hat{\theta}$ followed by a translation given by $\hat{x}$ and $\hat{y}$. Thus, evaluating (\ref{eq:space_representation}) in (\ref{eq:rotation}) a transformation of $\Gamma (x,y)\rightarrow \Gamma(x_l,y_l,\hat{x},\hat{y},\hat{\theta})$ can be reached. Where $x_l, y_l$ are the elements to integrate (surface under the curve) and $\hat{x},\hat{y},\hat{\theta}$ are the optimization variables (parking position of the car). In order to compute the integration we assume 1)~rectangular representations of the car body and its maneuvering areas and 2)~angle congruence between all the rectangles of the car, i.e.,~the rectangle of the maneuvering areas is composed of lines parallel to the car body.

The definition of the optimization algorithm is then defined by (\ref{eq:volume}).

\begin{equation}
  \label{eq:volume}
  \zeta(\hat{x},\hat{y},\hat{\theta}) = \sum_{s \in S} \int_{s_x}\int_{s_y} \Gamma(x_l,y_l,\hat{x},\hat{y},\hat{\theta})dx_ldy_l
\end{equation}

From~(\ref{eq:volume}), $S$ is the set of the rectangles that represent the car (car body and maneuvering areas) containing the properties $s_x$ and $s_y$ which is the interval of the rectangle on the $x$ and $y$ axes respectively. Finally, to find the position of the car the minimisation from equation (\ref{eq:optimization}) is computed.

\begin{equation}
  \label{eq:optimization}
  \begin{aligned}
& \underset{\hat{x},\hat{y},\hat{\theta}}{\text{minimize}}
& & \zeta(\hat{x},\hat{y},\hat{\theta}) \\
& \text{s.t.} & &  \hat{x} \geq 0, \quad \hat{x} \leq l_x \\
&  & &  \hat{y} \geq 0, \quad \hat{y} \leq l_y\\
\end{aligned}
\end{equation}

This is the minimization of (\ref{eq:volume}) subject to the limits of the parking space. The ideal constraint shall be that all the polygons representing the car should be inside the parking spot. However, this constraint can be relaxed using only $l_x$ and $l_y$ as position constraints.

To run the model in the computer, an approximation of (\ref{eq:volume}) was performed using a sampled function of (\ref{eq:space_representation}) over (\ref{eq:rotation}). This is needed because of the non-continuity caused by the $min$ and $max$ functions of (\ref{eq:space_representation}).

\subsection{Parking Strategy Application} \label{subsection:ParkingStrategyApplication}

When choosing from several parking spots, the vehicle picks the one that best minimizes constraints according to the optimization algorithm, with an alignment strategy. However, our prototype vehicle not being technically capable of positioning accurately, a simplified strategy is rounded-up based on the actual computation: for instance, to prioritize space on left or on right and on in-front or at back of the vehicle (see Figure~\ref{figure:hmi:proposition} for an example).

The test vehicle has been modified to receive these commands from CAN--Bus and execute them on our prototype \gls{ecu} (see Section~\ref{subsection:Materials}).

\section{Results}\label{section:Results}
\thispagestyle{empty}

\subsection{Algorithmic Results on Parking Strategy} \label{subsection:Results_ParkingStrategy}
\begin{figure}[!htb]
    \centering
    \subfloat[\centering Simplest scenario]{\includegraphics[width=.445\linewidth]{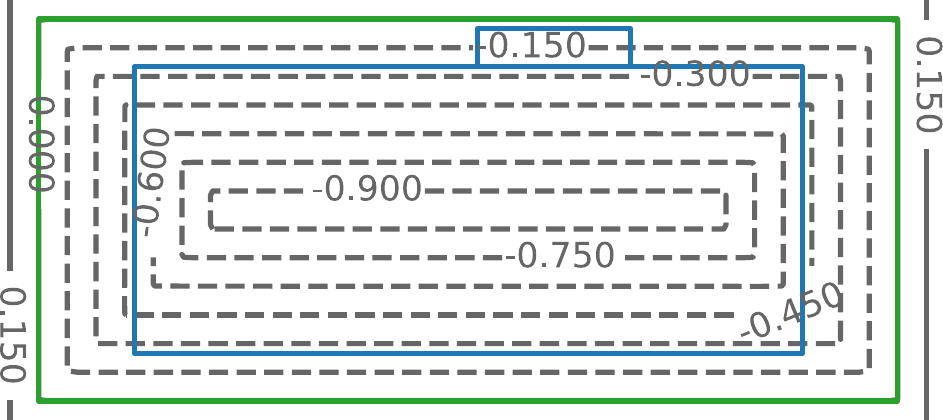}\label{fig:result_1} }
    \qquad
    \subfloat[\centering A scenario with a single obstacle.]{\includegraphics[width=.445\linewidth]{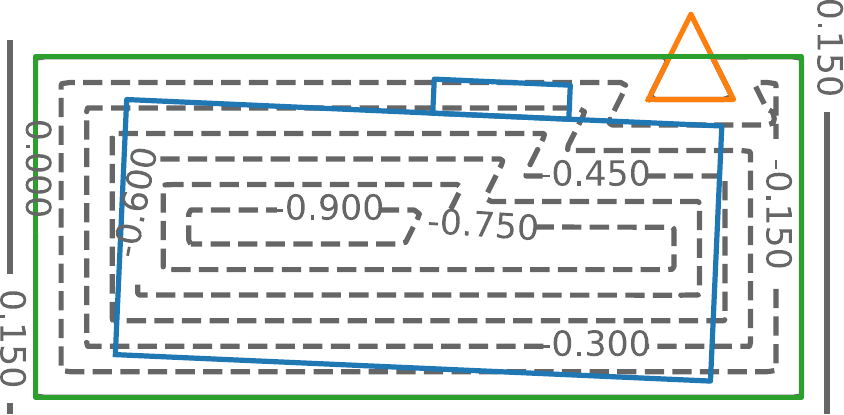}\label{fig:result_4} }
    \qquad
    \subfloat[\centering A scenario with two obstacles.]{\includegraphics[width=.445\linewidth]{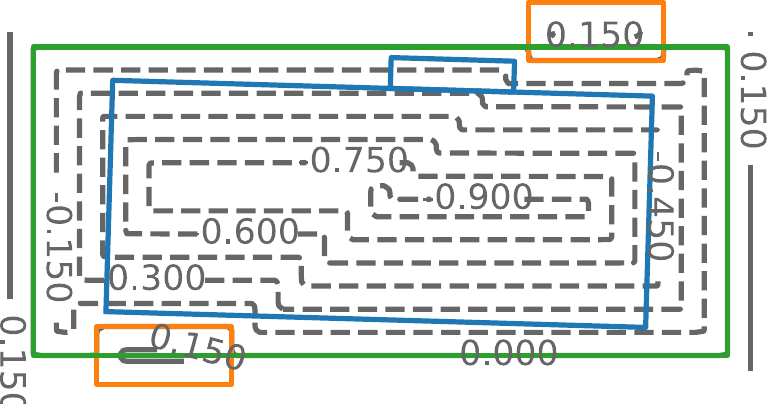}\label{fig:result_2} }
    \qquad    
    \subfloat[\centering A scenario with two different obstacles.]{\includegraphics[width=.445\linewidth]{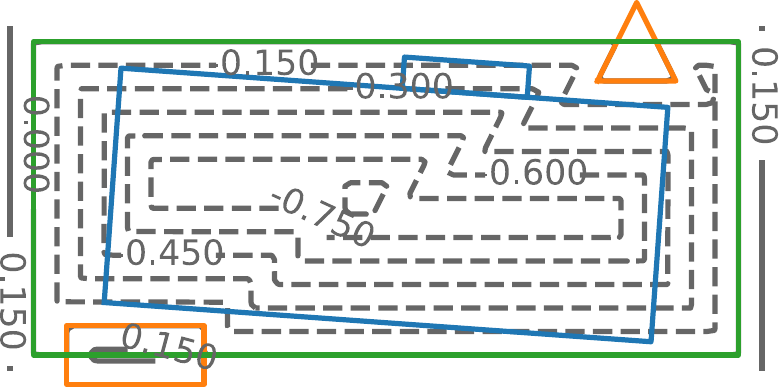}\label{fig:result_3} }
    \qquad    
    \subfloat[\centering A scenario with a single obstacle and with a baby seat.]{\includegraphics[width=.445\linewidth]{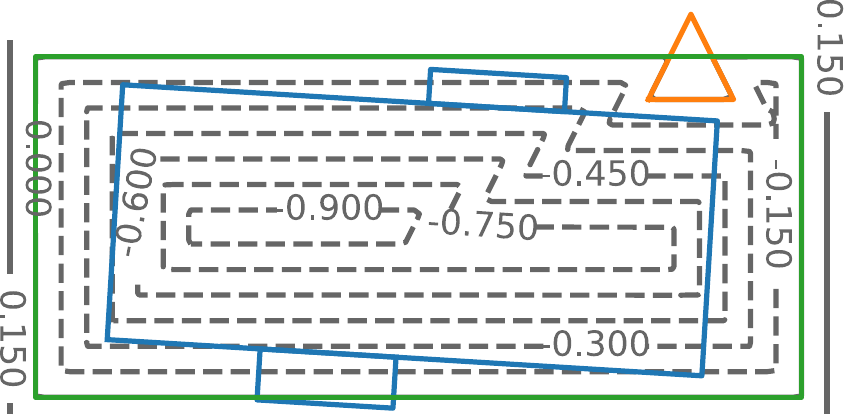}\label{fig:result_4_2} }
    \qquad
    \subfloat[\centering A scenario where another vehicle parked to the next parking spot.]{\includegraphics[width=.445\linewidth]{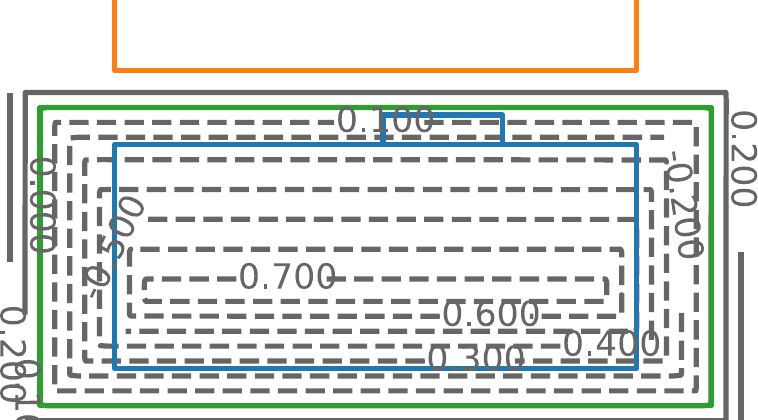}\label{fig:result_5} }
    \caption{Qualitative results of the parking algorithm}
    \label{fig:results}
\end{figure}

Subsection~\ref{subsection:AlgorithmicCore} explained how the optimization algorithm works. Polygons were used to represent the environment which generate force fields (see Figure~\ref{fig:force_field}) that are evaluated on the surface of the desired configuration of the car. Thus, Figures~\ref{fig:result_1}~to~\ref{fig:result_5} are qualitative results of our parking algorithm.

Figure~\ref{fig:result_1} depicts the result of the simplest scenario. The car is in an empty spot with no obstacles, every edge of the parking spot can be seen as a rectangle with height zero. The gray lines depict the force field and the car position (in blue) minimizes the surface over the area where the car stands. The car consists on two blue rectangles, one representing the car body and the second representing the maneuvering space for the driver. Figure~\ref{fig:result_4}~to~\ref{fig:result_3} depict a similar scenario as in Figure~\ref{fig:result_1}. However, there are some obstacles in the environment. Thus, the proposed approach finds an appropriate solution to the given scenarios. Figure~\ref{fig:result_4_2} shows the same environment as in Figure~\ref{fig:result_4}, however, a bigger maneuvering space is added to the car, due to the presence of a baby in the car. Finally, Figure~\ref{fig:result_5} is similar to Figures~\ref{fig:result_1}, however there is another car parked next the target parking spot. Gray lines indicate the deformation of the force field. Despite this deformation, the car is not so much pushed to the lower part of the picture. It can be assumed that this is due to the approximation function as representation of (\ref{eq:space_representation}). The approximation function is Monte Carlo integration with a few number of points (mainly in the edges of the polygons). 
Increasing the number of points sampled with a uniform distribution through all the area of the car and its safety areas could overcome this issue.

This system provides a parking strategy that can be used in a wide variety of scenarios.

\subsection{Pre-Test Interview} \label{results:pretest}
\thispagestyle{empty}

The pre-test interview (see Section~\ref{PreTestInterviews}), investigated users' general definitions of assistants as well as (manual) parking preferences and habits. The following quotes summarize the feedback collected:

\begin{itemize}
\item "\textit{[An assistant is] something that helps me. It reminds me to finish my tasks.}"
\end{itemize}
\begin{itemize}
\item "\textit{[An assistant is] something that helps me and simplifies things for me. It should be available when needed. It should not be annoying, and easy to use.}"
\item “\textit{It [the assistant] should somehow support me in my tasks, not doing all, but helping me.}"
\item "\textit{It [the assistant] organizes my life to be more easy.}"
\item "\textit{It [the assistant] gives recommendations. It should not force something. The user has to have an option to follow recommendation.}"
\end{itemize}

Thus, from users' perspectives, an assistant would be a support to make life easier and minimize physical as well as mental effort. However, it seems that users mainly expect on-demand support rather than autonomous behaviour, and they still want to have the choice to reject the assistant's help.

Although users look for parking spots sufficiently big and close enough to destination, most users expressed not having particular preferences. Still, some mentioned caring about the adaptation to passengers or other road users as illustrated by the next quotes:

\begin{itemize}
\item "\textit{I need to fit in. No special preference. If I have passengers, I need to make sure they can get out.}"
\item  "\textit{Passengers are taken into account when parking. I sometimes tell them to go out before I park in.}"
\item  "\textit{I prefer to park in a way to not let my dog jump directly to the street.. to avoid risky situations.}"
\item  "\textit{When my son was a baby, I took care of it to get out the baby seat easily.}"
\item "\textit{It should have enough space to other car so other drivers should also be able to park in.}" 
\item "\textit{If I can choose, and I have a kid in the back, I let more space to the right.}"
\item  "\textit{I pay attention to passengers so the doors can open or to reach my trunk when it is full.}"
\item "\textit{I care of my passengers to let them get out easily. Need to be careful with elderly people. They literally rip off the cars door when they get out. Same for children.}"
\item "\textit{I try to leave more space to drivers next to me so to let them enter easily. I don't want them to scratch my car.}"
\end{itemize}

In the end, users' parking habits and priorities can be summarized in terms of:

\begin{itemize}
\item Caring about drivers of cars parked next to the ego car, ensuring they have enough room to get in or out
\item Caring about passengers in the ego car (children and elderly people in particular), ensuring that they can get out easily and safely
\end{itemize}

The results of the pre-test interviews helped understand users' parking habits and assistance expectations , and indeed validated the relevance of a smart automated parking system.

\subsection{User Experience On Park4U Mate System}

\begin{table*}[!htb]
\caption{Positive and negative qualitative user feedback collected during test drives, with sample quotes for each section, pronounced from users.\label{table:uxResults}}
\begin{center}
\begin{tabular}{@{} p{.475\linewidth} | p{.475\linewidth} @{}}
\textbf{What users liked} & \textbf{What users didn't like} \\\toprule
\multicolumn{2}{c}{Communication style \& content} \\\hline
\begin{itemize}
\item Assistant sounds friendly and clear
\item Content of information provided is important (e.g. information on free parking spots, readiness for starting the parking process
\end{itemize}
\textit{"Clear spoken and good understandable communication"}.
&
\begin{itemize}
\item Assistant talked too much sometimes (too long sentences)
\end{itemize}
\textit{ "The programmed friendly language could take too long in everyday life (seem cumbersome), tighter sentences would be more practical"}.
\\\midrule
\multicolumn{2}{c}{Input \& output modalities of the assistant} \\ \hline

\begin{itemize}
\item Intuitive and easy way of  activating the parking function
\end{itemize}
\textit{"Great concept and easy operation and interaction", "Would be perfect to start assistance like that. I would do that [...] no buttons to press!"}
&
\begin{itemize}
\item Being limited on voice interaction only [...] selecting visualized parking spots on a touch interactive display would be a good addition
\item The assistant should also talk visually to avoid long sentences, e.g. in lights or pictures and icons on the display) 
\end{itemize}

\textit{"No possibility to touch screen to choose desired parking spot.", " I would give a clearer and shorter command here, possibly with sound or light.}
\\\midrule

\multicolumn{2}{c}{Decision-making process of the assistant} \\ \hline

\begin{itemize}
\item Decisions are made 'human--like' since it adapts the parking strategy for the good of passengers
\end{itemize}
\textit{"System makes good choices.", "The system aligns the parked vehicle based on the passages. Taking a small child into account is particularly helpful."}
&

\begin{itemize}
\item  Assistant patronizes driver, suggesting more suitable spots even after the driver selects one himself (e.g. driver selects a parking spot that makes it difficult for him to take out groceries)
\item Assistant doesn't know the drivers preferences: it should learn parking preferences and patterns and should adapt them depending on the location and time of parking
\end{itemize}
\textit{"It would be better if the car is not depending on its sensors but if it was a learning system that adapts automatically to parking habits and personal routing"}
\\\midrule

\multicolumn{2}{c}{Assistant's audio and visual feedback during the parking process} \\ \hline

\begin{itemize}
\item Assistant provides transparency during the parking process, informing the driver on when it is getting ready to start parking, when it is in position to start parking and when it is finished.
\end{itemize}
\textit{"I felt safe, not stressed. I saw what the system is doing. It explains me the positioning.", "Communicating the main milestones is important."}
&
\begin{itemize}
\item Assistant could give more precise feedback in terms of finally starting the parking maneuver after telling that it is in position for parking
\end{itemize}
\textit{"It should communicate when parking starts", "It didn't give information when it is starting to move forward."}
\\\midrule

\multicolumn{2}{c}{General thoughts on a 'talking car', independent from \textit{Park4U Mate}} \\ \hline
\multicolumn{2}{p{.95\linewidth}}{
\begin{itemize}
\item Futuristic and intuitive
\item An easy way of learning what the car is capable to do, increasing trust
\item Engaging to activate complex assistance functions (can be useful to activate also other assistance functions via voice interaction)
\item Enhancing safety since it supports to keep the eyes on the road and to monitor the vehicles environment while parking
\item It might have potential to ease interaction for drivers with physical constraints
\end{itemize}
}

\\\bottomrule
\end{tabular}
\end{center}
\end{table*}

\subsubsection{Qualitative User Feedback Collected During Test Drives}
Table~\ref{table:uxResults} classifies expert and novice users' feedback in four big categories: 1) communication style and content, 2) input and output modalities, 3) decision-making process of the assistant, 4) auditory and visual feedback during the maneuver.

\subsubsection{Quantitative Results - Post--Test (\gls{ueq})}
\label{subsection:Results_UserExperience}

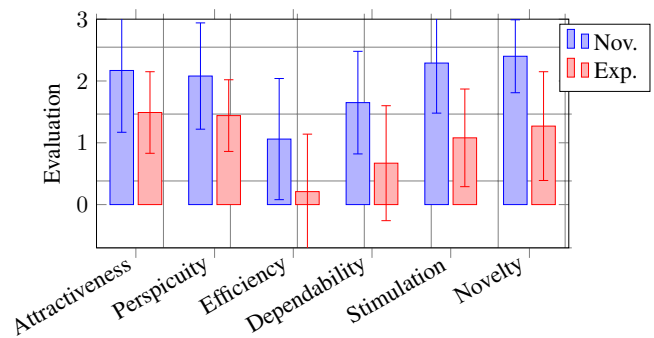
\begin{figure}[!htb]
 
    \centering
    \begin{tikzpicture} [scale=.89]
        \draw[step=1cm,gray,very thin] (0,0) grid (7.1,3.5);
        \begin{axis}[
            ybar,
            height=5cm,
            width=\linewidth,
            legend style={anchor=north west},
            ymin=-0.7,ymax=3,
            ylabel={Evaluation},
            symbolic x coords={Attractiveness,Perspicuity,Efficiency,Dependability,Stimulation,Novelty},
            xtick=data,
            x tick label style={rotate=30,anchor=east},
            ]
            \addplot+[error bars/.cd,y dir=both,y explicit] coordinates {
            (Attractiveness,2.17) +- (Attractiveness,1.0)
            (Perspicuity,2.08) +- (Perspicuity,0.86)
            (Efficiency,1.06) +- (Efficiency,0.98)
            (Dependability,1.65) +- (Dependability,0.83)
            (Stimulation,2.29) +- (Stimulation,0.81)
            (Novelty,2.40) +- (Novelty,0.59)
            };
            \addlegendentry{Nov.}
            \addplot+[error bars/.cd,y dir=both,y explicit] coordinates {
            (Attractiveness,1.49) +- (Attractiveness,0.66)
            (Perspicuity,1.44) +- (Perspicuity,0.58)
            (Efficiency,0.21) +- (Efficiency,0.93)
            (Dependability,0.67) +- (Dependability,0.93)
            (Stimulation,1.08) +- (Stimulation,0.79)
            (Novelty,1.27) +- (Novelty,0.88)
            };
            \addlegendentry{Exp.}
            \end{axis}
    \end{tikzpicture}

    \caption{Mean values for responses to the six \gls{ux} criterion, from twelve novice participants (in blue), and twelve expert participants (in red), where -3 is the worst score and +3 is the best score.}
    \label{table::UX_Questionnaire}
\end{figure}

As mentioned previously, for the quantitative study we received 12 questionnaire responses on the expert users and 12 on the novice users side (see \gls{ueq} in Section~\ref{EvaluationMethods}).
Figure~\ref{table::UX_Questionnaire} illustrates mean values for responses of the \gls{ueq}. It is noticeable that both user groups rated positively on each characteristic. However, novice users rated the global system better in every point of view compared to expert users. Both user groups rated similarly to the attractiveness and perspicuity but the difference of score on the other fields are more important. While novice users rated best to novelty and stimulation, experts rated best the attractiveness and perspicuity. The efficiency was the least rated criteria by both user groups. This derives from the fact that the system didn't meet the users' needs to perform tasks fast enough (without much talking) and users wish to have more options to initiate the parking maneuver (e.g. via touch display). Furthermore, the main reason for a significant gap between the user groups ratings within the dimension of "efficiency" is, that experts additionally also rated the parking performance. As experts in the field of automated parking, they have a much more critical view on the car's maneuvering and speed of parking.   

\unskip

\section{Conclusion}\label{section:Conclusion}
\thispagestyle{empty}

When designing an end-user feature of an automated system, we need to ponder the balance between intuitiveness, trustworthiness, understandability, and of course efficiency. In this work, we have looked at one common \gls{adas} function, the automated parking, and we have explored two enhancements: a perception of the situational context; and a voice-first human-machine interaction. We hypothesised these could help turn a primarily technical helper into a solution that feels natural for passengers to interact with, as well as aware of their needs. We evaluated the solution with automotive experts, knowledgeable about automated parking, and with novice users, not familiar with this \gls{adas} function. Test drives offered insights on users thoughts and feelings.

Results show that both user groups saw the benefits of such a system, capable of adapting its parking strategy to the context like a human driver does. Also, both user groups appreciated the ease of use of the voice assistant supported by the adaptive \gls{ui}. However, the feedback also told us that users would welcome less "talking" (i.e. more straight-to-the-point dialogues) and could use a more multi-modal interaction (i.e., touch gestures for quick selection). Interestingly, the collected quantitative data showed that expert users generally rated the system not as good as novice users. One major reason is, that experts in the field of automated parking expected a better, more efficient parking performance, whereas novice users generally were impressed by a "self-parking car". This, and another potential bias for our study is, the participants were only co-drivers. Their reactions and thoughts might be different if they were manipulating the system themselves.

Our work confirms that an advanced form of context awareness is a welcome expectation from end users when it comes to assisted driving functions. The algorithmic approach we proposed could be applied to many aspects of assisted and then automated driving (i.e., think about how a robo--taxi shall position itself when picking you up loaded with your luggage), complementing the technical solutions already implemented for safety and for automation, and possibly leveraging on the same set of sensors.

Additionally, our work confirms that, although voice interaction is a great helper for many attributes of  an \gls{adas} function (i.e, intuitiveness and understandability), the execution time becomes a users priority once the solution has been tried out and approved. As a next step, our solution could be further improved by a learning system, taking into account users previous experience and applying hidden automation, that is silently executing its purpose with little to no interaction, whenever possible.

\section*{Acknowledgments}\label{section:Acknowledgments}
\thispagestyle{empty}

The authors would like to thank: Katharina Hottelart for helping with user tests; Frédéric Gehin and Samuel Ravin for offering access to the \gls{ev}; Nicolas Jecker, Raoul Leumalieu-Pagop and Jeff Destrumelle for sharing their expertise in autonomous parking systems; Arnaud Huille and Alexandre Seferian for designing and implementing user interface; Pierre Escrieut and Uli Ritter for overseeing hardware modifications on the vehicle; Patrice Reilhac and Julien Moizard for supporting the project; and finally all experiment~participants for providing valuable feedback.

\bibliographystyle{./bibliography/IEEEtran}
\bibliography{./bibliography/bibliography}

\end{document}